\documentclass[twocolumn]{aastex631}

\shorttitle{First Metallicity Map from JWST}
\shortauthors{Wang et al. (2022)}

\hypersetup{linkcolor=dgreen,citecolor=lblue,filecolor=cyan,urlcolor=magenta}

\defcitealias{Roberts-Borsani.2022}{Paper I}

\usepackage[english]{babel} 
\usepackage[utf8]{inputenc} 
\usepackage[T1]{fontenc}    
\usepackage{ae,aecompl}
\usepackage{pgf,pgfarrows,pgfnodes,pgfautomata,pgfheaps}
\usepackage{graphicx}       
\usepackage{natbib}         
\usepackage{url}            
\usepackage{grffile}        
\usepackage{mathtools}      
\usepackage{multirow}       
\usepackage{xspace}         

\usepackage{amsmath,amssymb,amsxtra,amsfonts}   
\usepackage{txfonts}

\usepackage{color}         
\definecolor{gold}{rgb}{1,0.80,0}
\definecolor{orange}{rgb}{1,0.5,0}
\definecolor{midgray}{gray}{0.3}
\definecolor{lblue}{rgb}{0,0.2,0.6}
\definecolor{dgreen}{rgb}{0.1,0.6,0.3}
\definecolor{purple}{rgb}{0.5019607843137255,0.0,0.5019607843137255}

\renewcommand\farcs{\mbox{$.\!^{\prime\prime}$}}    
\renewcommand{\arcdeg}{\mbox{$^{\circ}$}\xspace}             

\newcommand{\be}{\begin{equation}}
\newcommand{\ee}{\end{equation}}

\newcommand{\ba}{\begin{align}}
\newcommand{\ea}{\end{align}}

\newcommand{\Msun}{\ensuremath{M_\odot}\xspace}

\newcommand{\Mstar}{\ensuremath{M_\ast}\xspace}

\newcommand{\Sstar}{\ensuremath{\Sigma_\ast}\xspace}
\newcommand{\oh}{\ensuremath{12+\log({\rm O/H})}\xspace}

\newcommand{\K}{\ensuremath{\rm K}\xspace}

\newcommand{\Hunit}{\ensuremath{\rm km~s^{-1}~Mpc^{-1}}\xspace}
\newcommand{\Funit}{\ensuremath{\rm erg~s^{-1}~cm^{-2}}\xspace}

\def\micron{\ensuremath{\mu\textrm{m}}\xspace}  

\newcommand{\Hb}{\textrm{H}\ensuremath{\beta}\xspace}
\newcommand{\Hg}{\textrm{H}\ensuremath{\gamma}\xspace}

\newcommand{\HII}{\textrm{H}\textsc{ii}\xspace}

\newcommand{\OII}{[\textrm{O}~\textsc{ii}]\xspace}
\newcommand{\OIII}{[\textrm{O}~\textsc{iii}]\xspace}

\newcommand{\NeIII}{[\textrm{Ne}~\textsc{iii}]\xspace}

\def\U{\ensuremath{U_{360}}\xspace}     
\def\B{\ensuremath{B_{435}}\xspace}
\def\V{\ensuremath{V_{606}}\xspace}
\def\I{\ensuremath{I_{814}}\xspace}
\def\Y{\ensuremath{Y_{105}}\xspace}
\def\J{\ensuremath{J_{125}}\xspace}

\def\H{\ensuremath{H_{160}}\xspace}

\newcommand{\grzl}{\textsc{Grizli}\xspace}

\newcommand{\bagp}{\textsc{BAGPIPES}\xspace}

\newcommand{\jwst}{\textit{JWST}\xspace}

\newcommand{\hst}{\textit{HST}\xspace}

\newcommand{\glass}{\textit{GLASS}\xspace}

\newcommand{\hff}{\textit{HFF}\xspace}

\newcommand{\mg}{\textit{MAMMOTH-Grism}\xspace}

\def\ie{i.e.\xspace}
\def\eg{e.g.\xspace}
\def\etc{etc.\xspace}

\renewcommand\({\left(}
\renewcommand\){\right)}

\newcommand{\el}[1]{\ensuremath{\textrm{EL}_{#1}}}

\def\p{{\rm prior}}

\newcommand{\n}{\ensuremath{{\nu}\rm}\xspace}

\usepackage{etoolbox}
\makeatletter
\patchcmd{\NAT@citex}
  {\@citea\NAT@hyper@{\NAT@nmfmt{\NAT@nm}\NAT@date}}
  {\@citea\NAT@nmfmt{\NAT@nm}\NAT@hyper@{\NAT@date}}
  {}
  {}
\patchcmd{\NAT@citex}
  {\@citea\NAT@hyper@{%
     \NAT@nmfmt{\NAT@nm}%
     \hyper@natlinkbreak{\NAT@aysep\NAT@spacechar}{\@citeb\@extra@b@citeb}%
     \NAT@date}}
  {\@citea\NAT@nmfmt{\NAT@nm}%
   \NAT@aysep\NAT@spacechar%
   \NAT@hyper@{\NAT@date}}
  {}
  {}
\patchcmd{\NAT@citex}
  {\@citea\NAT@hyper@{%
     \NAT@nmfmt{\NAT@nm}%
     \hyper@natlinkbreak{\NAT@spacechar\NAT@@open\if*#1*\else#1\NAT@spacechar\fi}%
       {\@citeb\@extra@b@citeb}%
     \NAT@date}}
  {\@citea\NAT@nmfmt{\NAT@nm}%
   \NAT@spacechar\NAT@@open\if*#1*\else#1\NAT@spacechar\fi%
   \NAT@hyper@{\NAT@date}}
  {}
  {}
\makeatother

\newcommand{\galname}{\glass-Zgrad1\xspace}

\begin{document}


\title{Early results from \glass-\jwst. IV: Spatially resolved metallicity in a low-mass $z\sim3$ galaxy with NIRISS\footnote{Based on observations acquired by the \jwst under the ERS program ID 1324 (PI T. Treu)}}

\correspondingauthor{Xin Wang}
\email{xwang@ucas.ac.cn}

\author[0000-0002-9373-3865]{Xin Wang}
\affil{School of Astronomy and Space Science, University of Chinese Academy of Sciences (UCAS), Beijing 100049, China}
\affil{National Astronomical Observatories, Chinese Academy of Sciences, Beijing 100101, China}
\affil{Infrared Processing and Analysis Center, Caltech, 1200 E. California Blvd., Pasadena, CA 91125, USA}

\author[0000-0001-5860-3419]{Tucker Jones}
\affiliation{Department of Physics and Astronomy, University of California Davis, 1 Shields Avenue, Davis, CA 95616, USA}

\author[0000-0003-0980-1499]{Benedetta Vulcani}
\affiliation{INAF Osservatorio Astronomico di Padova, vicolo dell'Osservatorio 5, 35122 Padova, Italy}

\author[0000-0002-8460-0390]{Tommaso Treu}
\affiliation{Department of Physics and Astronomy, University of California, Los Angeles, 430 Portola Plaza, Los Angeles, CA 90095, USA}

\author[0000-0002-8512-1404]{Takahiro Morishita}
\affil{Infrared Processing and Analysis Center, Caltech, 1200 E. California Blvd., Pasadena, CA 91125, USA}

\author[0000-0002-4140-1367]{Guido Roberts-Borsani}
\affiliation{Department of Physics and Astronomy, University of California, Los Angeles, 430 Portola Plaza, Los Angeles, CA 90095, USA}

\author[0000-0001-6919-1237]{Matthew A. Malkan}
\affiliation{Department of Physics and Astronomy, University of California, Los Angeles, 430 Portola Plaza, Los Angeles, CA 90095, USA}

\author[0000-0002-6586-4446]{Alaina Henry}
\affiliation{Space Telescope Science Institute, 3700 San Martin Drive, Baltimore MD, 21218} 
\affiliation{Center for Astrophysical Sciences, Department of Physics and Astronomy, Johns Hopkins University, Baltimore, MD, 21218}


\author[0000-0003-2680-005X]{Gabriel Brammer}
\affiliation{Cosmic Dawn Center (DAWN), Denmark}
\affiliation{Niels Bohr Institute, University of Copenhagen, Jagtvej 128, DK-2200 Copenhagen N, Denmark}

\author[0000-0002-6338-7295]{Victoria Strait}
\affiliation{Cosmic Dawn Center (DAWN), Denmark}
\affiliation{Niels Bohr Institute, University of Copenhagen, Jagtvej 128, DK-2200 Copenhagen N, Denmark}

\author[0000-0001-5984-0395]{Maru\v{s}a Brada\v{c}}
\affiliation{University of Ljubljana, Department of Mathematics and Physics, Jadranska ulica 19, SI-1000 Ljubljana, Slovenia}
\affiliation{Department of Physics and Astronomy, University of California Davis, 1 Shields Avenue, Davis, CA 95616, USA}


\author[0000-0003-4109-304X]{Kristan Boyett}
\affiliation{School of Physics, University of Melbourne, Parkville 3010, VIC, Australia}
\affiliation{ARC Centre of Excellence for All Sky Astrophysics in 3 Dimensions (ASTRO 3D), Australia}

\author[0000-0003-2536-1614]{Antonello Calabr\`o}
\affiliation{INAF Osservatorio Astronomico di Roma, Via Frascati 33, 00078 Monteporzio Catone, Rome, Italy}

\author[0000-0001-9875-8263]{Marco Castellano}
\affiliation{INAF Osservatorio Astronomico di Roma, Via Frascati 33, 00078 Monteporzio Catone, Rome, Italy}

\author[0000-0003-3820-2823]{Adriano Fontana}
\affiliation{INAF Osservatorio Astronomico di Roma, Via Frascati 33, 00078 Monteporzio Catone, Rome, Italy}

\author[0000-0002-3254-9044]{Karl Glazebrook}
\affiliation{Centre for Astrophysics and Supercomputing, Swinburne University of Technology, PO Box 218, Hawthorn, VIC 3122, Australia}

\author[0000-0003-3142-997X]{Patrick L. Kelly}
\affiliation{School of Physics and Astronomy, University of Minnesota, 116 Church Street SE, Minneapolis, MN 55455, USA}

\author[0000-0003-4570-3159]{Nicha Leethochawalit}
\affiliation{School of Physics, University of Melbourne, Parkville 3010, VIC, Australia}
\affiliation{ARC Centre of Excellence for All Sky Astrophysics in 3 Dimensions (ASTRO 3D), Australia}
\affiliation{National Astronomical Research Institute of Thailand (NARIT), Mae Rim, Chiang Mai, 50180, Thailand}

\author[0000-0001-9002-3502]{Danilo Marchesini}
\affiliation{Department of Physics and Astronomy, Tufts University, 574 Boston Ave., Medford, MA 02155, USA}

\author[0000-0002-9334-8705]{P. Santini}
\affiliation{INAF Osservatorio Astronomico di Roma, Via Frascati 33, 00078 Monteporzio Catone, Rome, Italy}

\author[0000-0001-9391-305X]{M. Trenti}
\affiliation{School of Physics, University of Melbourne, Parkville 3010, VIC, Australia}
\affiliation{ARC Centre of Excellence for All Sky Astrophysics in 3 Dimensions (ASTRO 3D), Australia}

\author[0000-0002-8434-880X]{Lilan Yang}
\affiliation{Kavli Institute for the Physics and Mathematics of the Universe, The University of Tokyo, Kashiwa, Japan 277-8583}


\begin{abstract}
We report the first gas-phase metallicity map of a distant galaxy measured with the James Webb Space Telescope (\jwst). We use the NIRISS slitless spectroscopy acquired by the \glass Early Release Science program to spatially resolve the rest-frame optical nebular emission lines in a gravitationally lensed galaxy at $z=3.06$ behind the Abell 2744 galaxy cluster.
This galaxy (dubbed \galname) has stellar mass $\sim10^{8.6}\Msun$, instantaneous star formation rate $\sim8.6$ \Msun/yr (both corrected for lensing magnification), and global metallicity one-fourth solar.
From its emission line maps (\OIII, \Hb, \Hg, \NeIII, and \OII) we derive its spatial distribution of gas-phase metallicity using a well-established forward-modeling Bayesian inference method.
The exquisite resolution and sensitivity of \jwst/NIRISS, combined with lensing magnification, enable us to resolve this $z\sim3$ dwarf galaxy in $\gtrsim$50 resolution elements with sufficient signal, an analysis hitherto not possible.
We find that the radial metallicity gradient of \galname is strongly inverted (\ie positive): $\Delta\log({\rm O/H})/\Delta r$ = $0.165\pm0.023$ $\mathrm{dex~kpc^{-1}}$.
This measurement is robust at $\gtrsim$4-$\sigma$ confidence level against known systematics.
This inverted gradient may be due to tidal torques induced by a massive nearby ($\sim$15 kpc projected) galaxy, which can cause inflows of metal-poor gas into the central regions of \galname.
These first results showcase the power of \jwst wide-field slitless spectroscopic modes to resolve the mass assembly and chemical enrichment of low-mass galaxies in and beyond the peak epoch of cosmic star formation ($z\gtrsim2$). Reaching masses $\lesssim 10^9~\Msun$ at these redshifts is especially valuable to constrain the effects of galactic feedback and environment, and is possible only with \jwst's new capabilities. 
\end{abstract}

\section{Introduction}\label{sect:intro}

A central challenge in galaxy evolution is to understand how galaxies assemble their baryonic mass, and cycle their gas and heavy elements in and around themselves.
Deep imaging and spectroscopic surveys with Hubble Space Telescope (\hst), the Sloan Digital Sky Survey (SDSS), and other facilities have provided a census of the cosmic star formation rate (SFR) and hence metal enrichment history, which peak at $1\lesssim z\lesssim3$ \citep[the ``cosmic noon'' epoch; \eg,][]{Madau.2014}.
Dedicated observational campaigns have established tight correlations among various physical properties of star-forming galaxies at this epoch, \eg, stellar mass (\Mstar), SFR, gas-phase metallicity, size, morphology, kinematics, and dust and gas content
\citep[see recent reviews by ][]{Kewley.2019,Maiolino.2019}.
These integrated scaling relations are largely reproduced by theoretical semi-analytic models and simulations which incorporate accretion and merging, star formation, and feedback in the form of gaseous outflows powered by star formation and supermassive black hole accretion \citep[\eg,][]{Somerville.2015}.
However, key properties of galactic outflows such as their mass loss rates and recycling timescales have proven extremely challenging to discern, especially at intermediate and high redshifts. A variety of assumptions and ``sub-grid'' physical models are able to reproduce global galaxy scaling relations. Hence, additional observational constraints are needed to better understand how feedback redistributes baryons within and around galaxies. 

Gas-phase metallicity gradients are sensitive probes of the complex gas flows driven by galactic feedback and tidal interactions.
The value of gas metallicity maps as a diagnostic of galaxy formation has been demonstrated by numerous integral field spectroscopy (IFS) surveys at $z=0$
\citep[\eg,][]{Sanchez.2014,Belfiore.2017,Poetrodjojo.2018,Franchetto.20212m}.
Theoretical models and simulations make differing predictions for metallicity gradient evolution, depending on the strength of feedback and the outflow properties at high redshifts \citep{Gibson.2013to}.
Predictions diverge the most at low stellar masses ($\Mstar$\,$\lesssim$\,$10^9\Msun$), especially at $z$>2 \citep{Mott.2013, Molla.2018, Gibson.2013to, Hemler.2021}. This is a key regime where feedback may be important for dynamical evolution and is even a candidate to resolve the longstanding ``cusp-core'' problem \citep[and related challenges to the cold dark matter paradigm; \eg,][]{Bullock.2016}. 
However, observations are challenging due to small angular sizes of high-$z$ galaxies, and ground-based adaptive optics studies have yielded limited samples \cite[\eg,][]{Jones.2013,Yuan.2013,Leethochawalit.2016, Curti.2020}. 
Meanwhile, \hst grism surveys have made substantial progress in charting resolved chemical enrichment at high redshifts \citep{Jones.2015,Wang.2017qqi,Wang.2020nq,Simons.2020gwa,Li.2022}. Combining WFC3 grism spectroscopy with gravitational lensing magnification, \citet{Wang.2020nq} assembled a large sample of galaxies with (sub-kiloparsec) sub-kpc resolution metallicity radial gradients at cosmic noon.
While this progress is encouraging, results from \hst are limited by wavelength coverage to $z\lesssim2.3$ and typically probe $\Mstar \gtrsim 10^9~\Msun$.

The Near-infrared Imager and Slitless Spectrograph (NIRISS; \citealt{Willott.2022a3}) onboard the James Webb Space Telescope (\jwst) now enables a tremendous leap forward with its superior sensitivity, angular resolution, and longer wavelength coverage compared to \hst/WFC3.
This allows resolved metallicity studies probing higher redshifts and lower mass galaxies, which are a powerful discriminant of feedback and outflow models \citep[\eg,][]{Ma.2017sec,Hemler.2021}.
This Letter presents the first study of gas-phase metallicity maps at $z>3$, a regime not yet explored observationally, based on the NIRISS data acquired by the Early Release Science (ERS) program \glass-\jwst \citep[ID ERS-1324\footnote{\url{https://www.stsci.edu/jwst/science-execution/approved-programs/dd-ers/program-1324}};][]{Treu.2022}. 
These observations also take advantage of the gravitational lensing magnification by the galaxy cluster Abell 2744, to study background high-$z$ sources in high definition.

This Letter is structured as follows. We briefly describe the observations in Section~\ref{sect:obs}. In Section~\ref{sect:measure} we describe the extraction of emission line maps and subsequent measurements of gas-phase metallicity (both the global value and spatial map). We discuss the results in Section~\ref{sect:results} and summarize the main conclusions in Section~\ref{sect:conclu}. 
We adopt a standard cosmology with $\Omega_{m}=0.3$, $\Omega_{\Lambda}=0.7$, and $\mathrm{H_0}=70$~\Hunit, and a \cite{Chabrier.2003} Initial Mass Function (IMF).

\section{Observations}\label{sect:obs}

In this Letter we use \jwst/NIRISS data from GLASS-ERS, whose observing strategy is described by \cite{Treu.2022}, and data reduction in \citet[Paper I][]{Roberts-Borsani.2022}. In brief, the Abell 2744 cluster core was observed with $\sim$15 hours with NIRISS wide field slitless spectroscopy and direct imaging in three filters (F115W, F150W, F200W). This provides low resolution $R\sim150$ spectra of all objects in the field of view with continuous wavelength coverage from $\lambda\in[1.0, 2.2]$~\micron\ (rest-frame [0.24, 0.54]~\micron\  at $z$=3.06). This includes the strong rest-frame optical emission lines [\textrm{Mg}~\textsc{ii}], 
\OII, \NeIII, \Hg, \Hb, and \OIII. Spectra are taken at two orthogonal dispersion angles (using the GR150C and GR150R grism elements) which helps to minimize the effects of contamination by overlapping spectral traces.

\section{Data analysis}\label{sect:measure}

\begin{figure*}
    \centering
    \includegraphics[width=.16\textwidth]{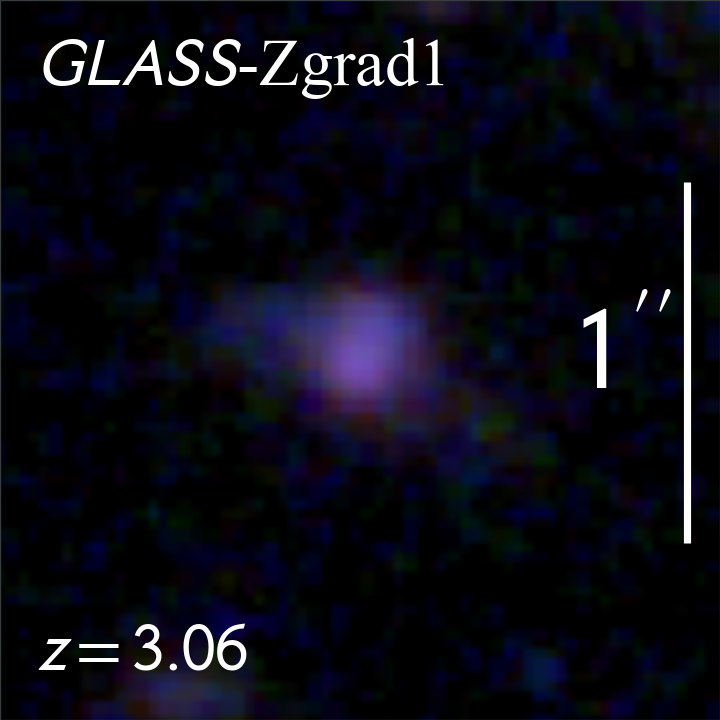}
    \includegraphics[width=.16\textwidth]{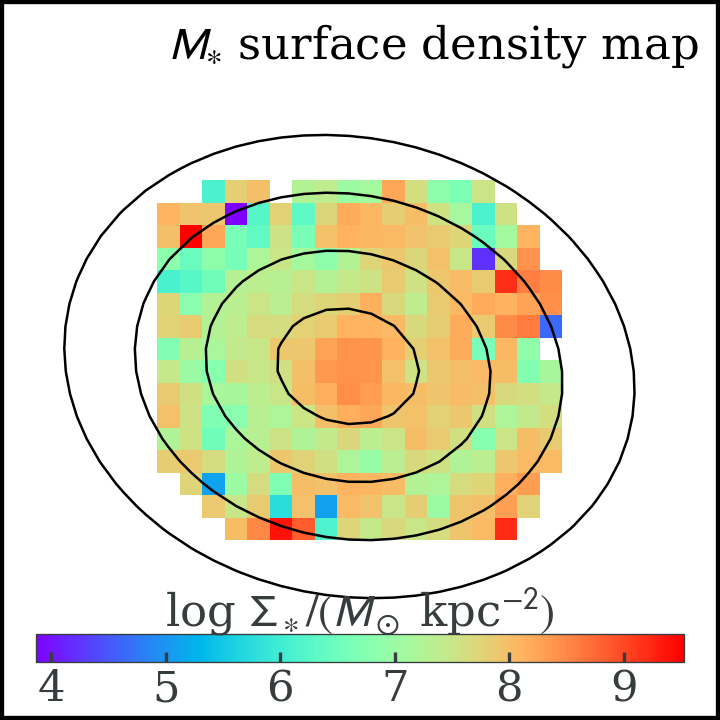}
    \includegraphics[width=.16\textwidth]{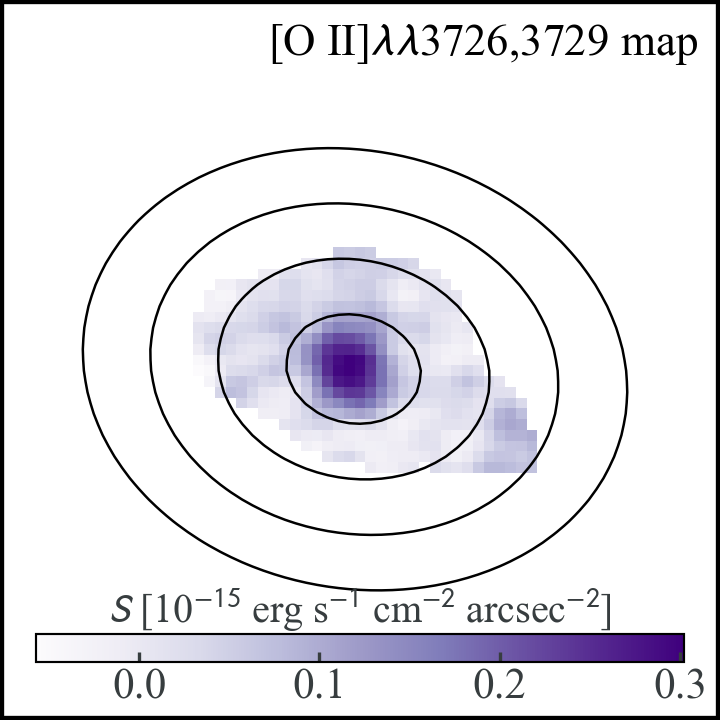}
    \includegraphics[width=.16\textwidth]{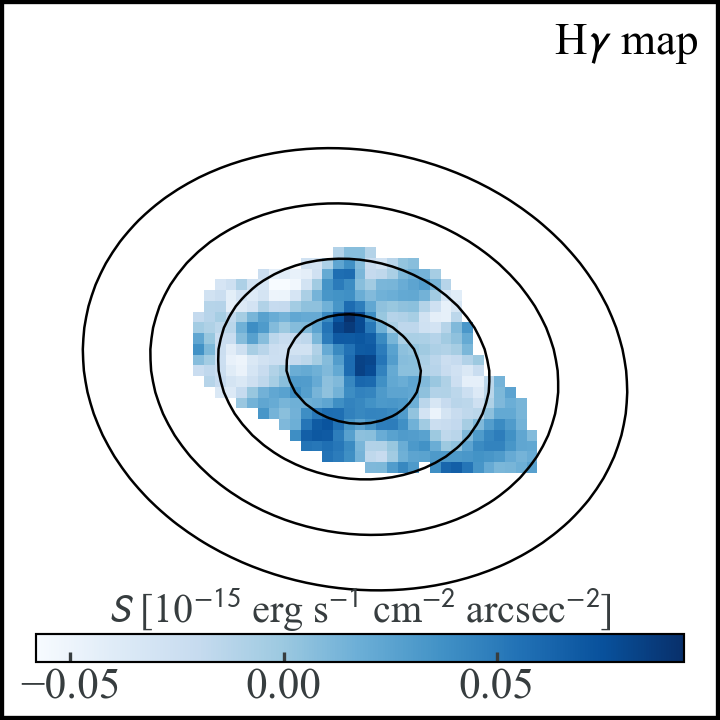}
    \includegraphics[width=.16\textwidth]{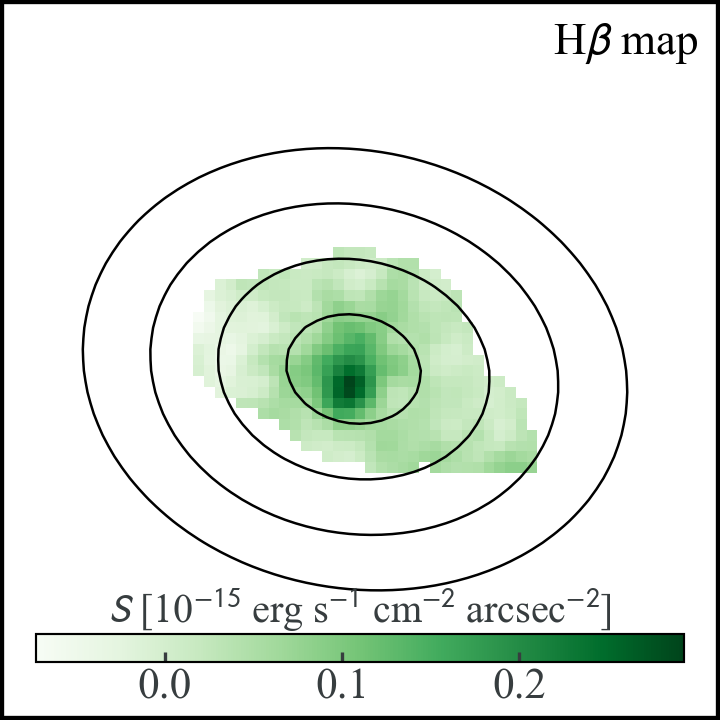}
    \includegraphics[width=.16\textwidth]{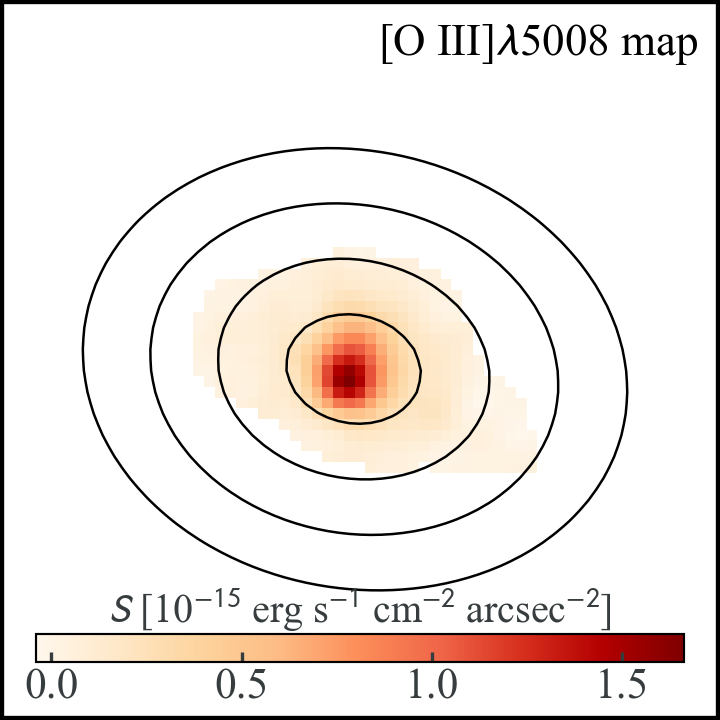}   \\
    \includegraphics[width=\textwidth]{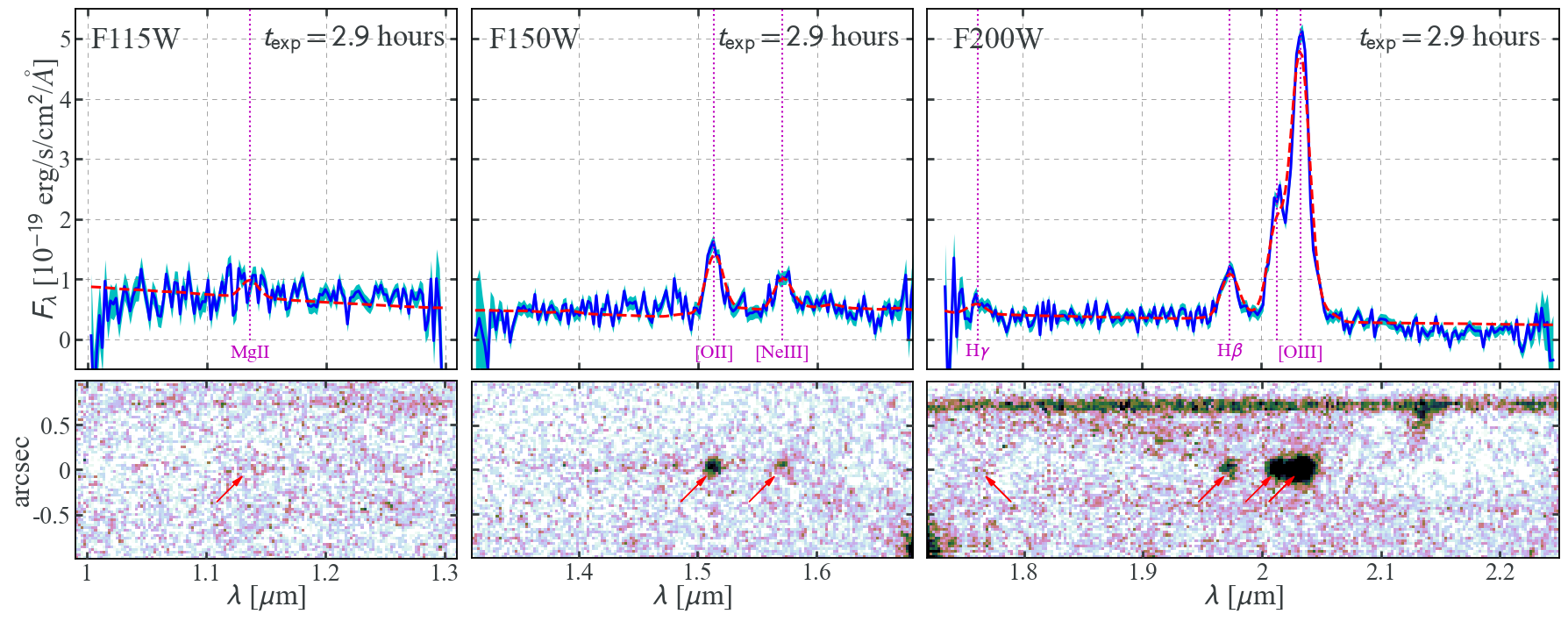}
    \vspace*{-2em}
    \caption{\label{fig:combEL}
    Galaxy \galname, a $z\sim$3 star-forming dwarf galaxy ($\Mstar$$\simeq$$10^{8.6}$\Msun) with a strongly inverted metallicity radial gradient presented in this work: the first metallicity map obtained with the \jwst and first ever such measurement secured by sufficient spatial resolution ($\lesssim$kpc) at $z\geq3$.
    {\bf Top, from left to right}: the color composite stamp (made using the 3-band \jwst/NIRISS pre-imaging on 30 mas plate scale), stellar surface density (\Sstar) map (derived from the
    pixel-by-pixel SED fitting), and surface brightness maps of emission lines (\OII, \Hg, \Hb and \OIII), measured from the NIRISS slitless spectroscopy acquired by \glass-\jwst.
    The black contours mark the de-lensed de-projected galacto-centric radii with 1 kpc interval, given by our source plane morphological reconstruction using the public lens model.
    The orientation (north up and east to the left) and spatial extent remain unchanged throughout all the 2D maps.
    {\bf Middle}: the optimally extracted 1D observed $F_{\lambda}$ flux and its 1-$\sigma$ uncertainty, represented by the blue solid lines and cyan shaded bands, respectively, combined from the two orthogonal dispersion directions (GR150C and GR150R). In total, the exposure time \emph{per} filter combined from both dispersions reaches 2.9 hours.
    The forward-modeled spectra (source continuum + nebular emission), based on the best-fit intrinsic SED with source morphological broadening taken into account, is shown in the red dashed line.
    {\bf Bottom}: the \emph{continuum-subtracted} 2D grism spectra covered by the three filters (F115W, F150W, and F200W) respectively, showcasing the quality of our \glass-\jwst slitless spectroscopy.
    }
\end{figure*}

Here we focus on the NIRISS observations of one galaxy (\galname) at $z=3.06$, whose measured properties are given in Table~\ref{tab:gal}.
\galname has highly secure grism redshift determination given by \grzl, with $\left(\Delta z\right)_{\rm posterior}/(1+z_{\rm peak})=1.7\times10^{-4}$, ascribed to its extremely strong rest-frame optical emission lines.
Its data benefit from low contamination from neighboring sources, so its spectrum is suitable to measure emission line maps.
The synergy of \jwst's resolving power and its lensing magnification \citep[estimated to be $\sim$3,][]{Johnson.2014,Wang.2015mhw,Bergamini.2022} reveals its detailed physical properties on scales as small as $\sim$200 pc. These characteristics make our target a textbook case to showcase the capabilities of \jwst in spatially resolving the properties of high-$z$ galaxies. 
The upper left panel of Fig.~\ref{fig:combEL} shows its color composite image, made by combining the three NIRISS filters (F115W, F150W and F200W).
The galaxy shows a smooth light profile, with no clear signs of clumpy structures.
The location of \galname on the mass-excitation diagram \citep{Juneau.2014,Coil.2015} indicates negligible emission line contamination from an active galactic nucleus (AGN), such that we can apply standard gas-phase metallicity diagnostics for \HII\ regions.

While \galname itself shows no clear signs of interaction, it is located only $\sim$15 kpc in projection from a $\sim100\times$ more massive galaxy ($\Mstar \approx 10^{10.6}~\Msun$) at a consistent redshift \citep{Wu.2022,Sun.2022}. This massive companion shows potentially spiral structure which may be induced by interaction with \galname \citep{Wu.2022}. Such interaction may affect the metallicity gradient via torques on the gas; we discuss this possibility in Section~\ref{sect:results}.

\begin{table}
\centering
\begin{tabular}{lc}
\hline
Galaxy & \galname \\
\hline
R.A. [deg] & 3.585943 \\
Dec  [deg] & -30.382102 \\
$z_{\rm spec}$     & 3.06 \\
$\mu$   & $3.00^{+0.03}_{-0.03}$  \\
\hline
\multicolumn{2}{c}{Observed emission line fluxes}   \\
$f_{\OIII}$~[$10^{-19}$\Funit]     &  913.77 $\pm$ 9.65   \\
$f_{\Hb}$~[$10^{-19}$\Funit]       &  161.68 $\pm$ 8.89  \\
$f_{\Hg}$~[$10^{-19}$\Funit]       &   40.71 $\pm$ 12.73   \\
$f_{\NeIII}$~[$10^{-19}$\Funit]      &  101.65 $\pm$ 14.76  \\
$f_{\OII}$~[$10^{-19}$\Funit]      &  207.43 $\pm$ 9.65  \\
\hline
\multicolumn{2}{c}{Rest-frame equivalent widths}    \\
EW$_{\OIII}$~[\AA]     &  749.37 $\pm$ 25.23   \\
EW$_{\Hb}$~[\AA]       &  156.92 $\pm$ 10.18  \\
EW$_{\Hg}$~[\AA]       &   31.66 $\pm$ 9.39  \\
EW$_{\NeIII}$~[\AA]    &   47.80 $\pm$ 6.51  \\
EW$_{\OII}$~[\AA]      &  130.52 $\pm$ 6.70  \\
\hline
\multicolumn{2}{c}{Nebular emission diagnostics}   \\
\oh$_{\rm global}$ &  $8.11^{+0.07}_{-0.06}$    \\
$\Delta\log({\rm O/H})/\Delta r$ [$\mathrm{dex~kpc^{-1}}$] & $0.165\pm0.023$ \\
SFR$^{\rm N}$  [$M_\sun~{\rm yr}^{-1}$] &  $8.64^{+6.35}_{-2.48}$\\
log(sSFR$^{\rm N}$/[${\rm yr}^{-1}$]) &  $-7.68^{+0.34}_{-0.22}$\\
$A_V^{\rm N}$ & $0.54^{+0.63}_{-0.39}$  \\
\hline
\multicolumn{2}{c}{Broad-band photometry SED fitting}   \\
$\log(M_\ast/M_\sun)$ & $8.62^{+0.07}_{-0.10}$ \\
SFR$^{\rm S}$  [$M_\sun~{\rm yr}^{-1}$] &  $4.28^{+0.43}_{-0.44}$\\
log(sSFR$^{\rm S}$/[${\rm yr}^{-1}$]) &  $-7.97^{+0.07}_{-0.10}$\\
$A_V^{\rm S}$ & $0.60^{+0.03}_{-0.04}$  \\
\hline
\end{tabular}
\caption{Physical properties of galaxy \galname. The values of \Mstar and SFR have been corrected for lensing magnification.}
\label{tab:gal}
\end{table}

\subsection{Emission line map and metallicity measurements}

We rely on the latest version the Grism Redshift \& Line software \citep[\grzl\footnote{\url{https://github.com/gbrammer/grizli}};][]{2021zndo...5012699B} to extract and model the grism spectra, following the procedure described in \citetalias{Roberts-Borsani.2022}. \grzl provides contamination subtracted 1D and 2D grism spectra, along with the best-fit spectroscopic redshifts, as shown in Fig.~\ref{fig:combEL} \citep[also see Appendix A of][for the full descriptions of the redshift fitting procedure employed by \grzl]{Wang.2019czf}. 
We follow the procedures described by \citet{Wang.2020nq} to extract un-contaminated line maps from the 2D grism spectra, also shown in Fig.~\ref{fig:combEL}. 
The maps shown here are encompassed by the boundary of the \grzl segmentation map \citep[for details see][]{Roberts-Borsani.2022} defined using conventional source detection parameters largely following the 3D-HST survey \citep{Brammer.2012,Momcheva.2016lbw}. We restrict our analysis to this region, where the stellar continuum is detected at high significance.
We additionally use the spatially integrated line fluxes measured by \grzl to estimate global physical properties as presented in Table~\ref{tab:gal}. When drizzling the 2D grism spectra and emission line maps, we adopt a plate scale of 30 mas, to Nyquist sample the FWHM of the \jwst point spread function at these wavelengths.

We jointly constrain ionized gas metallicity (\oh), nebular dust extinction ($A_v^{\rm N}$), and de-reddened \Hb line flux ($f_{\Hb}$), using our forward-modeling Bayesian inference method, largely described in \cite{Wang.2017qqi,Wang.2019czf,Wang.2020nq,Wang.2022}. The likelihood function is defined as 
\begin{align}\label{eq:chi2}
    \mathrm{L}\propto\exp\left(-\frac{1}{2}\cdot\sum_i \frac{\(f_{\el{i}} - R_i \cdot 
    f_{\Hb}\)^2}{\(\sigma_{\el{i}}\)^2 + \(f_{\Hb}\)^2\cdot\(\sigma_{R_i}\)^2}\right).
\end{align}
where $f_{\el{i}}$ and $\sigma_{\el{i}}$ represent the de-reddened emission-line (\eg \OII, \Hg, \Hb, \OIII) flux and its uncertainty (measured from the flux-calibrated error array; see \citealt{Roberts-Borsani.2022}), corrected using the \citet{Calzetti.2000} dust extinction law with $A_v^{\rm N}$ as a free parameter.
$R_i = \frac{f_{\el{i}}}{f_{\Hb}}$ is the expected flux ratio of each line with respect to \Hb, based on the adopted metallicity calibrations and intrinsic Balmer decrements, while $\sigma_{R_i}$ is the intrinsic scatter in $R_i$.
To compute metallicity we adopt the purely empirical strong line diagnostics prescribed by \citet{Bian.2018}, calibrated against a sample of local analogs of high-$z$ galaxies according to their location on the BPT \citep{Baldwin.1981} diagram.
We note that dust extinction $A_v^{\rm N}$ is constrained by the metallicity diagnostics in addition to H Balmer line ratios, as described in our earlier work. For example, the characteristic locus of R$_{23}$ versus O$_{32}$ \citep[see, e.g.,][]{Shapley.2015} and the intrinsic \OIII/\NeIII\ line ratios \citep{Jones.2015.Te} provide constraints on $A_v^{\rm N}$ along with traditional diagnostics such as \Hb/\Hg. The effect of uncertainty in $A_v^{\rm N}$ is included in our metallicity uncertainties.
To estimate the instantaneous star formation rate (SFR$^{\rm N}$), we use the Balmer line luminosity assuming the \citet{Kennicutt.19986u} calibration scaled to the \citet{Chabrier.2003} IMF, \ie, ${\rm SFR^N}=1.6\times 10^{-42} \frac{L(\Hb)}{\rm [erg ~s^{-1}]}~[M_\odot {~\rm yr^{-1}}]$. 

The forward-modeling Bayesian inference described above is first performed on the integrated emission line fluxes to yield global values of \oh, $A_v^{\rm N}$, and SFR$^{\rm N}$ (with results listed in Table~\ref{tab:gal}). We then utilize Voronoi tessellation \citep{Cappellari.2003,Diehl.2006} as in \cite{Wang.2019czf, Wang.2020nq} to construct spatial bins with nearly uniform signal-to-noise ratio (SNR) of \OIII, the strongest emission line available. We apply the Bayesian inference with line fluxes in each Voronoi bin, yielding maps of metallicity and other properties at sub-kpc resolution. Overall, we are able to resolve this $z\sim3$ dwarf galaxy with $\gtrsim$50 resolution elements at a signal-noise ratio threshold of 10 in \OIII.

As this work relies on emission lines spanning multiple spectroscopic filters, we have examined the relative flux calibration accuracy as a potential source of error. Comparing the spectra with both broadband photometric colors and best-fit spectral templates, we estimate that the relative calibration is accurate to within $\lesssim$10\% near the center of each filter (F115W, F150W, and F200W). This applies to the strong lines of \OII, \NeIII, \Hb, and \OIII in our $z=3.06$ target. Calibration uncertainty for \NeIII/\OII and \OIII/\Hb appears to be negligible. 
However, fluxes near band edges appear to be underestimated in some cases, possibly due to slight misalignment of the expected and actual spectral trace positions. This potentially affects the \Hg line such that its true flux may be $\sim$20\% larger than we report here. 
We note that systematic error of this magnitude is smaller than the statistical uncertainty for spatial Voronoi bins (where line ratio uncertainties are $\gtrsim$10\% and \Hg is undetected). Any relative calibration error between the F150W and F200W spectra would appear similar to a change in $A_v$; in this case our methodology would produce a bias in $A_v$ but not in \oh. Absolute flux calibration is not needed since our results rely entirely on line ratios. 
Flux calibration errors therefore may bias the derived $A_v$ but should not substantially affect the spatial trends in metallicity. While more detailed examination is beyond the scope of this work, for now we caution that modest NIRISS flux calibration uncertainties may exist and that wavelengths near the bandpass edges should be treated carefully. We anticipate that calibrations will improve over time, as the calibration plan is executed during Cycle-1.

\subsection{SED fitting and stellar mass map}\label{sec:mass_map}

There has been a wealth of deep multi-wavelength imaging in broad bands covering 0.2--5~\micron wavelengths in the field of A2744. We rely on the photometric catalog compiled by the ASTRODEEP project \citep{Merlin.2016}, which performs aperture-matched and blending-corrected photometry of the \hff 7-band, Hawk-I Ks, and Spitzer IRAC images \citep{Brammer.2016csj,Lotz.2017}. We use the \bagp software \citep{Carnall.2018} to fit the BC03 models \citep{Bruzual.2003} of galaxy spectral energy distributions (SEDs) to the ASTRODEEP photometry.
\bagp is capable of adding the nebular emission component into galaxy spectra, so the emission line fluxes are taken into account in the fit simultaneously.
Basic assumptions and parameter ranges include the \citet{Chabrier.2003} IMF, a stellar metallicity range of $Z/Z_{\odot}\in(0, 2.5)$, the \citet{Calzetti.2000} dust extinction law with $A_v^{\rm S}$ in the range of (0, 3), and an exponentially declining star formation history with $\tau$ in the range of (0.01, 10)\,Gyr.
We fix the spectroscopic redshift of our galaxy to its best-fit grism value ($z=3.06$), with a conservative uncertainty of $\Delta z/(1+z)\approx 0.003$, following \citet{Momcheva.2016lbw}. The obtained physical properties are presented in Table~\ref{tab:gal}. The measured stellar mass and SFR$^{\rm S}$ are $\log(\Mstar/\Msun)=8.62^{+0.07}_{-0.10}$ and ${\rm SFR^{S}}=4.28^{+0.43}_{-0.44}$, respectively, indicating that our galaxy lies 0.14 dex above the star-forming main sequence \citep{Speagle.2014}.
We also note that our derived values of the dust reddening for the nebular and stellar component are consistent with E(B-V)$_{\rm stars}$ $\sim$ E(B-V)$_{\rm gas}$ \citep{Reddy.2015}.

Broad-band image morphology is not necessarily a good tracer of the underlying stellar mass distribution in star forming galaxies \citep[\eg,][]{Wuyts.2012}. We, therefore, construct stellar mass maps using spatially resolved SED fitting. We use the same method as in \cite{Wang.2020nq}, applied to Hubble Frontier Field imaging \citep{Lotz.2017}. 
The F606W, F814W, F105W, F125W, F140W, and F160W HST images are used for this analysis. F435W is not used as it corresponds to the Lyman $\alpha$ forest region at $z=3.06$.
Images from each filter are aligned and convolved to a common angular resolution of 0\farcs22 FWHM, ensuring good sensitivity at low surface brightness reaching beyond the effective radius. The photometry in each 0\farcs06-square pixel is then fit using FAST \citep{Kriek.2009} with an exponentially declining star formation history, \cite{Chabrier.2003} IMF, and \cite{Bruzual.2003} stellar population synthesis models \citep[for full details see][]{Wang.2020nq}. The resulting stellar mass map is shown in Fig. \ref{fig:combEL}. We fit the stellar mass map with a 2-D elliptical Gaussian in order to determine the galactocentric radius at each point \citep{Jones.2015}. 
The stellar mass surface density is indeed smoother than the luminosity distribution, and we consider morphological parameters based on this mass map to be more reliable. 

Future \jwst photometry with NIRCam, especially in filters relatively unaffected by strong nebular emission, will enable improved SED fitting results via spatially resolved photometry at longer wavelengths. Our spatial SED fitting is currently based on HST photometry sampling only $\lesssim4100$~\AA\ in the rest frame. 
NIRISS F200W imaging would extend the wavelength coverage, but it contains strong emission lines which account for $\sim$55\% of the total broad-band flux based on the \Hb\ and \OIII\ equivalent widths. While this can be corrected using the emission line maps (Figure~\ref{fig:combEL}), the effect of including this additional photometry is small and comparable to the uncertainty. We therefore do not use it for this initial study. 

\subsection{Source plane morphology}

\begin{figure*}[t!]
    \centering
    \includegraphics[width=\textwidth,trim=0cm 0cm 0cm 0cm,clip]{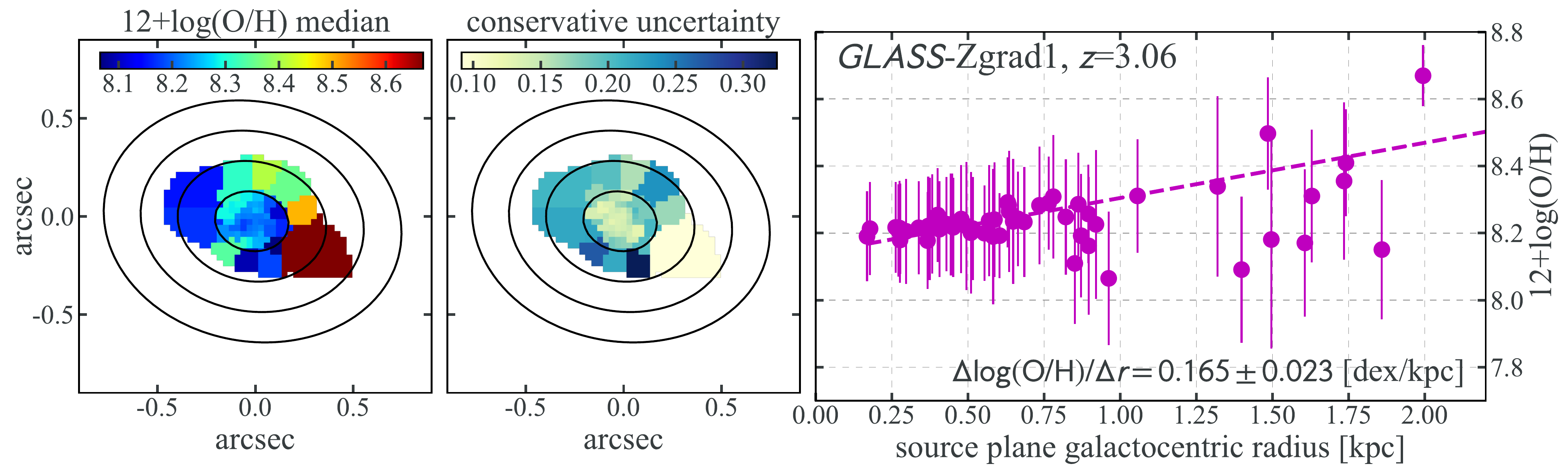}
    \vspace*{-2em}
    \caption{\label{fig:metalmap}
    The metallicity spatial maps and radial gradient of \galname at $z\sim3$.
    In the left and central panels, we adopt the weighted Voronoi tessellation method \citep{Cappellari.2003,Diehl.2006} to divide the galaxy surface into spatial bins with a uniform SNR of 10 on \OIII. The black contours represent the delensed, deprojected galactocentric radii with 1 kpc intervals from our source-plane morphological reconstruction.
    In the right panel, we show the radial gradient of the metallicity inferences in all individial Voronoi cells, with a slope of $\Delta\log({\rm O/H})/\Delta r= 0.165\pm0.023~[\mathrm{dex~kpc^{-1}}]$ represented by the magenta dashed line.
    }
\end{figure*}

We perform a source plane reconstruction of \galname to recover its intrinsic morphology based on public lens models \citep{Johnson.2014,Wang.2015mhw,Bergamini.2022}. Each pixel in the image-plane stellar mass map is ray-traced back to its source-plane position based on the deflection fields from the macroscopic cluster lens model. We then fit a 2D Gaussian profile to galaxy's intrinsic stellar mass spatial map to pinpoint its axis ratio, orientation and inclination, so that we obtain the intrinsic source morphology (corrected for lensing distortion). 
The centroid of this 2D fit is adopted as the galaxy center ($r=0$).
As a result, we establish the delensed deprojected galactocentric distance scale for each Voronoi bin, as displayed by the black contours in Figs.~\ref{fig:combEL} and \ref{fig:metalmap}.

\section{The first metallicity map from \jwst/NIRISS wide-field slitless spectroscopy}\label{sect:results}

Here we present the main results of our analysis. The top panels of Fig.~\ref{fig:combEL} show the maps of the most prominent emission lines of \galname: \OII, \Hg, \Hb, \OIII. Its optimally extracted 1D spectrum, covered by all three grism filters (F115W, F150W and F200W) combined from both dispersion elements (GR150C and GR150R), is shown in the central row, while the \emph{continuum-subtracted} 2D grism spectrum is shown in the bottom row. All nebular emission features used in this work are clearly seen. This 1D/2D spectrum showcases the quality of our NIRISS spatially resolved spectroscopy of high-$z$ emission line galaxies.

The emission line maps unveil how most of the emission is concentrated in the galaxy core, sharply decreasing at radii $>$1 kpc. In the outskirts the emission looks quite smooth, especially for \OIII. 
The only exception is \Hg, which appears patchy due to the lower SNR of this line.

Fig.~\ref{fig:metalmap} shows the 2D metallicity map of \galname. 
We note that variations in the derived metallicity result directly from the intrinsic emission line flux ratios via Equation~\ref{eq:chi2} (e.g., as shown for the R$_{23}$ and O$_{32}$ indices for two examples in Figure~2 of \citealt{Wang.2019czf}).
Clearly, the galaxy outskirts on average display highly elevated metallicities, \ie, more metal-enriched by $\sim$0.2 dex than the center. To quantify the radial gradient, we conduct a linear regression using a simple least-squares method, with the functional form $\oh=\theta_1 r + \theta_0$. Here $\theta_1$ and $\theta_0$ are the radial gradient and the intercept, respectively, and $r$ is the source-plane galactocentric radius in kpc. We obtain the best-fit and 1-$\sigma$ uncertainty results as $\theta_1 \equiv \Delta\log({\rm O/H})/\Delta r = 0.165\pm0.023~ [\mathrm{dex~kpc^{-1}}]$ and $\theta_0 = 8.140\pm0.019$. The central metallicity is roughly one-third of the solar value \citep[adopting solar \oh=8.69;][]{Asplund.2009} with a significant positive radial gradient. 
The positive metallicity gradient in \galname is even steeper than those measured by \cite{Wang.2019czf} for two similarly low-mass $z\sim2$ galaxies, which also show extremely high emission line equivalent widths, and which have gradients of $\sim 0.1$ dex/kpc. 

We performed several tests to verify that the positive gradient slope is robust against potential systematic errors. 
Firstly, we have binned the data in radial annuli to obtain a gradient measurement which is independent of the Voronoi tesselation. Using radial bin sizes such that each has SNR~$>10$ in \OIII, we find $\theta_1 = 0.14\pm0.03$~dex~kpc$^{-1}$ which is fully consistent with our fiducial result \citep[see also similar tests in][]{Wang.2017qqi}. 
Secondly, we consider whether spurious noise features could bias the results. The largest possible effect is from the highest-metallicity region which is also at the largest radius ($r\simeq2$~kpc). Repeating the analysis with this point removed still results in a significant positive gradient $\theta_1 = 0.076\pm0.019$~dex~kpc$^{-1}$, albeit smaller than our fiducial result by design. 
Thirdly, we vary the galaxy center (i.e., $r=0$ position) by up to 0\farcs1 and find no significant difference in the results. 
Finally, to test the statistical significance of the positive slope, we also calculated the Pearson correlation coefficient $r_P$ for metallicity as a function of radius. The result is $r_P = 0.43$ (p-value 0.001) which confirms the significance of the positive slope independent of statistical uncertainty estimates. 
Taken together, we conclude from these tests that the metallicity gradient slope is indeed positive and on the order of $\gtrsim$0.1~dex~kpc$^{-1}$ at $\gtrsim4\sigma$ significance.

In addition to the steeply rising metallicities indicated by the strong and highly significant ($\geq$7-$\sigma$) inverted radial gradient in \galname, there also exist some hints of azimuthal metallicity variations. In particular the local metallicity shows large variations at radii $r\simeq1$--2 kpc. This azimuthal structure and deviation from a smooth radial gradient, if further confirmed in numerous high-$z$ galaxies, can shed light upon the timescales for chemical enrichment relative to diffusion and mixing via measurements of the metallicity correlation length scale. This would represent a valuable new constraint for models of galaxy formation at high redshifts \citep[\eg,][]{Metha.2021,Metha.2022}.

\begin{figure*}[t!]
    \centering
    \includegraphics[width=.9\textwidth,trim=0cm 1.3cm 0cm 1.2cm,clip]{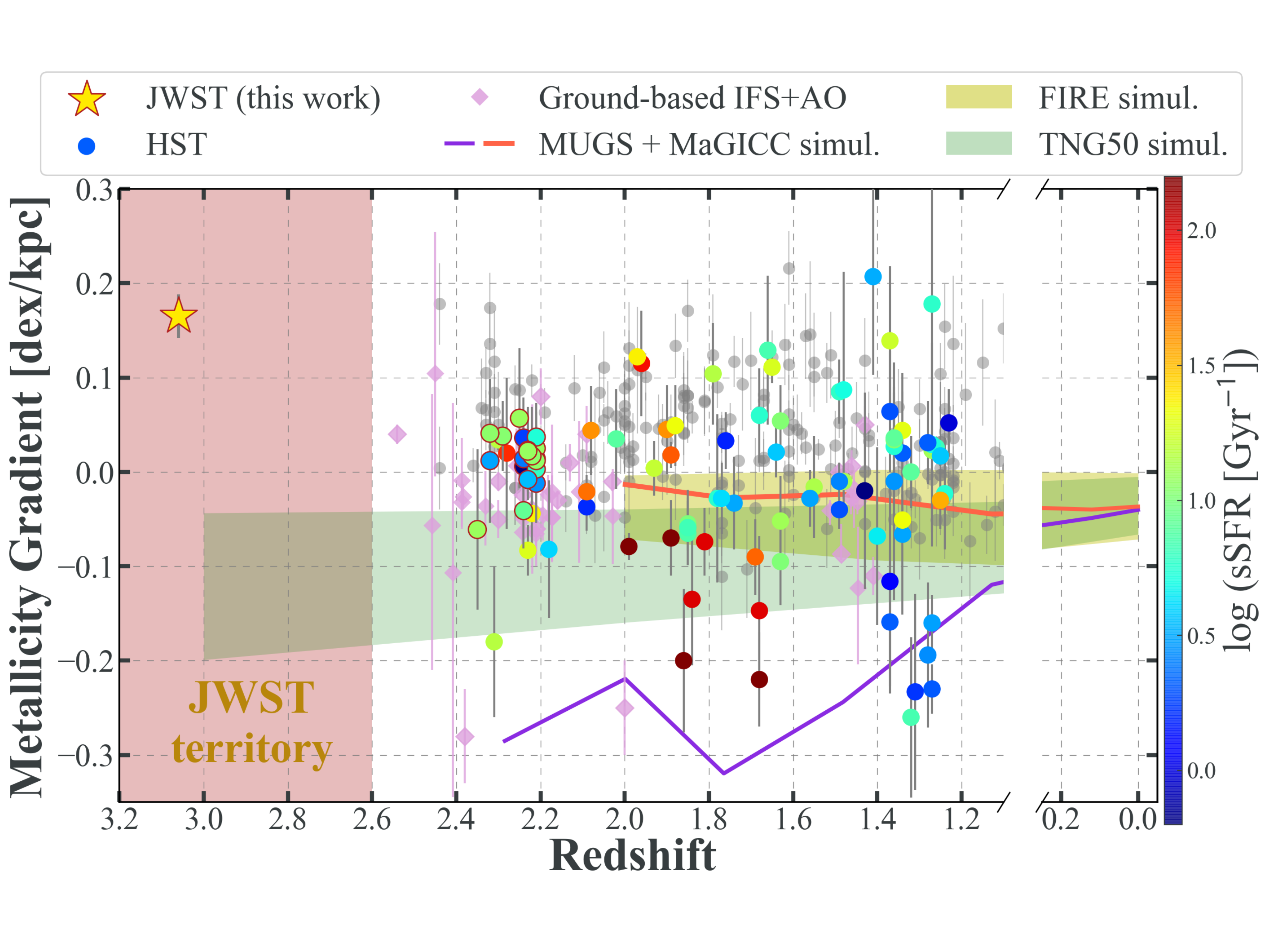}
    \vspace*{-1em}
    \caption{\label{fig:gradVSz}
    The redshift evolution of sub-kpc resolution metallicity gradients measured from observations and predicted by hydrodynamic simulations.
    Using the deep NIRISS slitless spectroscopy acquired by \glass-\jwst, we present hitherto the first metallicity gradient measurement with sub-kpc resolution at $z\gtrsim3$, highlighted by the star color-coded in sSFR.
    Before this work, all previous observational efforts at similar resolution ($\lesssim$kpc scale) primarily come from ground-based AO-assisted integral-field spectroscopy \citep[marked by the magenta diamonds,][]{Swinbank.2012,Jones.2013,Leethochawalit.2016,Schreiber.2018pdp} and space-based slitless spectroscopy with \hst/WFC3 \citep[marked by the circles also color-coded in sSFR if such information is publicly available,][]{Jones.2015,Wang.2020nq,Simons.2020gwa,Li.2022}.
    In particular, we highlight the first measurements of metallicity radial gradients 
    in high-$z$ overdense environments conducted by the \mg survey \citep[\hst-GO-16276, P.I. Wang,][]{Li.2022} using color-coded circles with maroon edges.
    In comparison, we show the predictions made by state-of-the-art cosmological hydrodynamic simulations: the 1-$\sigma$ spread of FIRE \citep{Ma.2017sec} and Illustris-TNG50 \citep{Hemler.2021}, as well as two Milky Way analogs with normal and enhanced feedback strengths but otherwise identical numerical setups \citep[corresponding to MUGS and MaGICC, respectively,][]{Gibson.2013to}.
    Despite the wide differences among the numerical recipes implemented, these simulations make indistinguishable predictions at $z\lesssim1$. However, their predictions diverge more significantly at higher redshifts, leaving a potential breakthrough at $z\gtrsim2.6$, where the \jwst spatially resolved spectroscopy can be highly efficient in deriving metallicity gradients and thus test these theoretical predictions.
    }
\end{figure*}

In Fig.~\ref{fig:gradVSz}, we show the redshift evolution of radial metallicity gradients in high-$z$ galaxies secured with sub-kpc spatial resolution.
This sub-kpc resolution is crucial in ensuring the results are robust against beam-smearing effects \citep[\eg,][]{Yuan.2013}.
The vast majority of the galaxies shown here reside in blank fields, with the exceptions of those identified by the \mg survey \citep[\hst-GO-16276, P.I. Wang,][]{Wang.2022} residing in the core regions of one of the most massive galaxy protoclusters at $z\sim2.24$ \citep[marked by the maroon edges,][]{Li.2022}.
The current existing sample in the literature consists of essentially two types of measurements driven by the available infrared spectroscopic instrumentation: ground-based IFS assisted with adaptive optics (AO) \citep{Swinbank.2012,Jones.2013,Leethochawalit.2016,Schreiber.2018pdp} and space-based slitless spectroscopy \citep{Jones.2015,Wang.2020nq,Simons.2020gwa,Li.2022}. The latter method only refers to \hst WFC3 spectroscopy before this work. Limited by the wavelength coverage of the WFC3/G141 grism and the H/K-band AO Strehl ratios, there have not yet been \emph{any} sub-kpc resolution measurements of metallicity maps at $z\geq3$. 
Using the deep NIRISS slitless spectroscopy acquired by \glass-\jwst, we for the first time secure an inverted gradient in the redshift range of $z\gtrsim2.6$.
This opens up a new key window on accurate characterization of the chemical profiles of galaxies in and beyond the cosmic noon epoch, to constrain the effect of galactic feedback.
Notably, we can see from Fig.~\ref{fig:gradVSz} that the theoretical predictions by a number of state-of-the-art cosmological hydrodynamic simulations --- FIRE \citep{Ma.2017sec}, Illustris-TNG50 \citep{Hemler.2021}, MUGS and MaGICC \citep{Gibson.2013to} --- significantly diverge at high redshifts, as a consequence of diverse feedback prescriptions.
They also struggle to reproduce the fraction of inverted (positive slope) gradients currently observed.

The existence of a relatively steep positive metallicity gradient in \galname is at first surprising, given the predominantly flat or moderately negative radial gradients in Fig.~\ref{fig:gradVSz}. However, \galname has a massive companion which may be gravitationally interacting (Section~\ref{sect:measure}; \citealt{Wu.2022}). Such an interaction can induce torques on the gas, causing metal-poor gas at large radii to lose angular momentum and migrate toward the center of the galaxy. This process can result in flattened or inverted metallicity gradients \citep[e.g.,][]{Kewley.2006}. 
Indeed, close gravitational interaction is the primary driver of positive gradients in $z\simeq0$ galaxies \citep[e.g.,][]{Rupke.2010,Torrey.2012}. We therefore consider it likely that the gradient in our target is indeed steeply positive and is induced by interaction with the nearby object.

\begin{figure*}[t!]
    \centering
    \includegraphics[width=.9\textwidth,trim=0cm 1.3cm 0cm 1.2cm,clip]{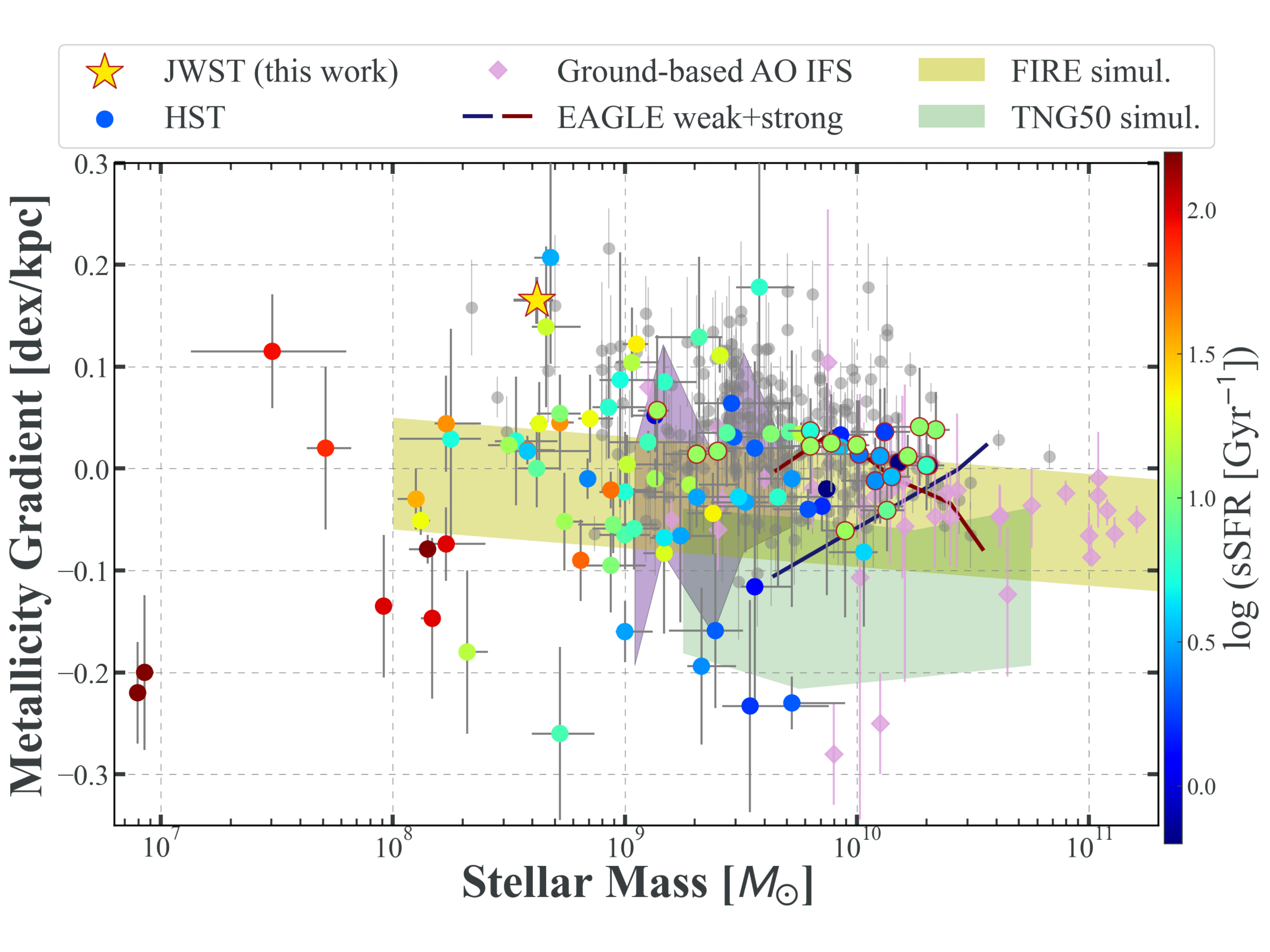}
    \vspace*{-1em}
    \caption{\label{fig:gradVSm}
    The mass dependence of sub-kpc resolution metallicity gradients measured from observations and predicted by hydrodynamic simulations. All symbols remain the same as in Fig.~\ref{fig:gradVSz}, except that the solid lines now represent the predictions by the EAGLE simulations assuming weak or strong feedback prescriptions \citep{Tissera.2019}. The $z=2$ results of the Illustris-TNG50 simulations are shown here. The latest EAGLE results extending to $\Mstar\sim10^9\Msun$ at $z=2$ are denoted by the purple shaded regions bracketing the [25th, 75th] percentiles of the simulation predictions, which shows a better agreement with the observed scatters in metallicity gradients
    \citep{Tissera.2021}.
    As compared to the \mg metallicity gradients measured in overdense environment having a sample median of $0.014\pm0.022~[\mathrm{dex~kpc^{-1}}]$ \citep{Li.2022}, 
    \galname shows a much more pronounced inverted gradient, indicating stronger environmental effects.
    This is consistent with the theoretical prediction that tidal torques, stemming from close gravitational interaction, induce cold gas infalls into the inner disk to invert metallicity radial gradient \citep{Torrey.2012,Jones.2013,Jones.2015}.
    }
\end{figure*}

In Fig.~\ref{fig:gradVSm}, we show the metallicity gradients of these field galaxies in terms of their \Mstar. This correlation is highly sensitive to galactic feedback. For instance, the two suites of EAGLE simulations using different feedback prescriptions predict widely dissimilar mass dependence \citep{Tissera.2019}. Observationally there is considerable scatter in the metallicity gradient slopes, which the current state-of-the-art numerical hydrodynamic simulations \citep[\eg][]{Ma.2017sec,Hemler.2021} struggle to reproduce \citep[but see][]{Tissera.2021}.

Recently, there is growing evidence for the existence of inverted metallicity gradients in distant galaxies \citep{Cresci.2010,Wang.2019czf,Li.2022}. \citet{Wang.2019czf} interpreted this phenomenon as an indication of the metal-enriched gas outflows triggered by central starbursts that disrupt galaxy out of equilibrium \citep[also see][]{Sharda.2021}, whereas other groups have suggested centrally directed low-metallicity gas infall as the physical cause \citep{Cresci.2010,Mott.2013}.
The latter scenario has been adopted to explain the inverted gradients observed in protocluster environments where gas inflows are easily drawn by the underlying large gravitational potentials \citep{Li.2022}.
Here our analysis confirmed a third scenario where strongly inverted gradients at high redshifts can be caused by the inflow of metal-poor gas into inner disks as a result of close gravitational interactions.
Furthermore, as shown in Figs.~\ref{fig:gradVSz} and \ref{fig:gradVSm}, the magnitude of the positive radial slope measured in \galname is significantly more dramatic than the sample median of $0.014\pm0.022~[\mathrm{dex~kpc^{-1}}]$ measured in the \mg galaxies residing in extreme overdensities \citep{Li.2022}.
This likely suggests that gravitational interactions impose stronger environmental effects on the chemical profiles of high-$z$ galaxies than large-scale overdensities.

\section{Conclusions}\label{sect:conclu}

In this Letter, we present the first spatially resolved analysis of a high-$z$ emission line galaxy using the wide-field slitless spectroscopic modes provided by \jwst.
We first showcase the quality of our NIRISS grism data acquired by the \glass-\jwst ERS program, presenting the extracted and cleaned 1D/2D grism spectra of a $z\simeq3$ dwarf galaxy, \galname.
Via SED fitting to the broad-band photometry fully covering its rest-frame UV through optical spectrum, we obtain $\log(\Mstar/\Msun)=8.62^{+0.07}_{-0.10}$ and ${\rm SFR^S} = 4.28^{+0.43}_{-0.44}~[M_\sun~{\rm yr}^{-1}]$ corrected for lensing magnification ($\mu=3.00{\pm}0.03$), $\sim$0.14 dex above the SFMS at $z\sim3$.
From the optimally extracted 1D grism spectrum (in particular the rest-frame optical nebular emission lines: \OIII, \Hb, \Hg, and \OII), we measure its global physical properties to be $\oh_{\rm global}=8.11^{+0.07}_{-0.06}$, ${\rm SFR^N} = 8.64^{+6.35}_{-2.48}~[M_\sun~{\rm yr}^{-1}]$, and $A_V^{\rm N} = 0.54^{+0.63}_{-0.39}$, consistent with E(B-V)$_{\rm stars}$ $\sim$ E(B-V)$_{\rm gas}$.

Importantly, the new capabilities afforded by \jwst/NIRISS (\eg exquisite sensitivity, resolution, and wavelength coverage, \etc), coupled with gravitation lensing magnification, enable us to spatially resolve \galname into $\gtrsim$50 individual resolution elements with source-plane resolution reaching $\sim$200 pc scales. Using its multiple nebular emission line maps, we determine its metallicity radial gradient to be highly inverted (positive), \ie, $\Delta\log({\rm O/H})/\Delta r$ = $0.165\pm0.023$ [$\mathrm{dex~kpc^{-1}}$]. This is the first-ever metallicity map from \jwst spectroscopy, and the first sub-kpc resolution metallicity gradient ever measured at $z\gtrsim3$. 
We also verified this positive gradient slope through extensive tests. In particular, by removing the outermost spatial bin at $r\simeq2$~kpc showing significant metal enhancement, we rederived the gradient slope to be $0.076\pm0.019$~dex~kpc$^{-1}$, vindicating our finding that in \galname the abundance of metals increases with galacto-centric radius.

This strongly inverted metallicity gradient of \galname likely stems from the close gravitational interaction with a nearby object --- a dusty galaxy $\sim100\times$ more massive than \galname at a projected separation of $\sim$15 kpc.
This close encounter results in powerful tidal torques to induce metal-poor gas inflows to the inner regions of \galname, inverting its radial metallicity gradient.
We thereby confirm a third channel for high-$z$ galaxies showing strongly inverted metallicity gradient, in addition to cold-mode accretion and metal-enriched gas outflows.
Moreover, in light of the more pronounced inverted gradient measured in \galname than any of the \mg gradient sources, we deduce that close gravitational interactions are more capable of producing strong environmental effects that modulate the chemical profiles of galaxies than large-scale overdensities (\eg protocluster cores) at high redshifts.
Our subsequent analysis of a larger sample within the \glass-\jwst data set will enable more comprehensive investigations of the true chemical profiles of galaxies at $z\sim3$, so that we can make a quantitative comparison to simulations and hence cast more meaningful constraints on the effects of mergers, galactic feedback, and environments. This work marks the beginning of a new era for spatially resolved analysis on galaxy chemo-structural evolution in and beyond the cosmic noon epoch.

\begin{acknowledgements}
We would like to thank the anonymous referee for careful reading and constructive comments that help improve the clarity of this paper.
This work is based on observations made with the NASA/ESA/CSA James Webb Space Telescope. The data were obtained from the Mikulski Archive for Space Telescopes at the Space Telescope Science Institute, which is operated by the Association of Universities for Research in Astronomy, Inc., under NASA contract NAS 5-03127 for JWST. These observations are associated with program JWST-ERS-1324. We acknowledge financial support from NASA through grant JWST-ERS-1324.
XW is supported by CAS Project for Young Scientists in Basic Research, Grant No. YSBR-062.
XW thanks Zach Hemler for kindly providing his compilations of gradient measurements in the literature, and Patricia Tissera for providing the tabulated results of the latest EAGLE simulations.
KG acknowledges support from Australian Research Council Laureate Fellowship FL180100060.  MB acknowledges support from the Slovenian national research agency ARRS through grant N1-0238. 

\end{acknowledgements}


\bibliography{reference}{}
\bibliographystyle{aasjournal}

\end{document}